# A Portable Optical Frequency Comb with Low Repetition-Rate Drift Based on “Optical Cubes”

Zhendong Chen, Hanbo Xu, Ruoao Yang, Tianyu Liu, Hongshuo Wu, Duo Pan, Zhigang Zhang, Jianjun Wu and Jingbiao Chen

***Abstract*—We demonstrate a compact, fully stabilized, all-polarization-maintaining (all-PM) Er:fiber optical frequency comb (OFC) implemented on a silica optical-cube platform. The optical comb module, including the oscillator, amplifier, spectral-broadening stage, 780-nm frequency-doubling branch, and *f*-to-2*f* interferometer, occupies a volume of 0.88 L. Without active temperature control, the free-running repetition frequency exhibits a peak-to-peak excursion of 75 Hz over 10 h, corresponding to 0.375 ppm. Under the same laboratory conditions, this excursion is 6.2 times smaller than that of a mechanically mounted non-PM reference comb. The amplified pulses are compressed to 94 fs and spectrally broadened in a PM highly nonlinear fiber. A coherent-population-trapping-based chip-scale atomic clock is employed for the first time as the radio- frequency reference for comb locking, thereby reducing the footprint of the microwave-reference subsystem. The resulting in-loop, optical-frequency-normalized tracking instability reaches $6.03\times10^{-17}$ at 1 s and $2.57\times10^{-18}$ at 10,000 s. These results demonstrate a compact comb architecture with reduced sensitivity to environmental perturbations, providing a practical route toward portable optical atomic clocks.**

***Index Terms*—Optical frequency comb; Coherent Population Trapping; Optical cubes; The repetition-rate drift.**

Manuscript received ×××; revised ×××; accepted ×××. Date of publication ×××; date of current version×××. This work was supported by the National Natural Science Foundation of China (12503074), the Beijing Nova Program (No.20240484696), the Wenzhou Major Science and Technology Innovation Key Project (No.ZG2023021), and the INNOVATION Program for Quantum Science and Technology (No.2021ZD0303202), Tianjin Key Laboratory of Quantum Precision Measurement Technology (2J-QT-J-25-003-SK). (Corresponding author: Ruoao Yang, Jianjun Wu)

Zhendong Chen is with the Institute of Quantum Electronics, School of Electronics, Peking University, Beijing 100871, China, the Tianjin Key Laboratory of Quantum Precision Measurement Technology, Tianjin 300131, China, and the School of Electronic and Information Engineering, Lanzhou Jiaotong University, Lanzhou 730070, China. (e-mail: 2201111540@stu.pku.edu.cn)

Hanbo Xu, Ruoao, Yang, Duo Pan, and Zhigang Zhang are with the Institute of Quantum Electronics, School of Electronics, Peking University, Beijing 100871, China (e-mail: 2501112263@stu.pku.edu.cn; ruoao.yang@pku.edu.cn; panduo@pku.edu.cn; jbchen@pku.edu.cn).

Hongshuo Wu are with the Tianjin Key Laboratory of Quantum Precision Measurement Technology, Tianjin 300131, China (e-mail: wuhongshuo@tjnavi.com).

Tianyu Liu is with China Quantum Measurement Tek Co., Ltd., Huzhou 313098, China (e-mail: Ltd.liu_ty@pku.edu.cn).

Jianjun Wu is with the Institute of Information and Communication, School of Electronics, Peking University, Beijing 100871, China (e-mail: just@pku.edu.cn).

Jingbiao Chen is with the Institute of Quantum Electronics, School of Electronics, Peking University, Beijing 100871, China, the National Key Laboratory of Advanced Micro and Nano Manufacture Technology, Beijing 100871, China, and Hefei National Laboratory, Hefei 230088, China (e-mail: jbchen@pku.edu.cn).

## I. Introduction

OPTICAL frequency combs (OFCs) generate a set of mutually coherent spectral lines with precisely defined frequency spacing, enabling phase-coherent links among the optical, microwave, and terahertz domains [1]. This capability has established OFCs as key tools for optical atomic clocks [2], [3], broadband precision spectroscopy [4], ultralow-noise microwave generation [5], and precision distance metrology [6]. As optical clocks evolve from laboratory instruments toward transportable and field-deployable systems, their frequency-comb subsystems must simultaneously provide low drift, environmental robustness, and a compact footprint. Recent field demonstrations, particularly in maritime environments, have further highlighted the need for portable and reliable OFCs [7], [8].

Among the most promising candidates for portable optical frequency references are thermal atomic- and molecular-vapor systems based on iodine molecules and alkali-metal vapor cells. Under optimized conditions, these systems can achieve short-term fractional frequency instabilities at the $10^{-14}$ level and approach $10^{-15}$ at longer averaging times. In 2024, NIST operated an iodine-based optical clock aboard a naval vessel for more than 20 days, achieving a daily timing error below 300 ps within a 35 L package [8]. Since 2015, Peking University has developed a series of compact optical frequency references at 420, 459, 780, 795, 852, and 1529 nm using modulation-transfer spectroscopy [9], [10], [11], [12], [13], [14], [15], [16], [17]. In such systems, an OFC provides the phase-coherent interface required to connect the optical reference to electronically accessible microwave outputs. Consequently, reducing the size of the OFC while improving its long-term and environmental stability is essential for fully integrated portable optical-clock systems [18], [19].

Conventional OFCs are assembled with metallic mechanical mounts on metal substrates, which gives rise to two major issues [19], [20], [21], [22], [23]. First, bulky metallic mounts limit the miniaturization of OFCs. Second, mechanical stress relaxation tends to induce optical path deviation. Meanwhile, metal substrates feature a high thermal expansion coefficient and are highly susceptible to temperature fluctuations, thereby degrading system stability. In 2021, Feng et al. mounted the optical components of a mode-locked solid-state laser directly on an aluminum baseplate [24]. Subsequently, Yang et al. demonstrated an integrated Yb-doped fiber laser by bonding free-space optical components onto silica substrates using an “optical-cube” architecture [25], and later extended this approach to an RF-referenced OFC [26]. However, the latter system employed nonlinear-polarization-evolution mode locking and non-polarization-maintaining fibers, making its operating state sensitive to polarization perturbations.

The nonlinear amplifying loop mirror (NALM) serves as a passive mode-locking mechanism that permits an all-polarization-maintaining (PM) design [27], [28]. Benefiting from the single-polarization propagation inherent to PM fibers, NALM mode-locked fiber lasers are well positioned for high environmental stability [29], [30], [31], [32]. Zhao et al. reported an Er-doped fiber laser with a nonlinear amplifying loop mirror on the "optical cubes" platform and confirmed its low timing jitter via the balanced optical cross-correlator technique. This approach provides a promising route toward compact and mechanically robust OFC systems [33].

In addition to the optical architecture, a compact and stable microwave reference is required for the RF synthesizers, phase-locking electronics, and frequency counters used in a fully stabilized OFC. Commercial hydrogen masers and rubidium frequency standards can provide suitable long-term frequency stability, but their size and power consumption hinder their integration into portable comb systems [27], [34]. Crystal oscillators are considerably more compact and can be directly integrated into the electronic subsystem [35]; however, their limited long-term stability may increase the long-term drift of the radio-frequency (RF) locking setpoints. Coherent-population-trapping (CPT)-based chip-scale atomic clocks (CSACs) combine a small footprint and low power consumption with improved long-term frequency stability relative to compact crystal oscillators, making them attractive microwave references for portable OFC electronics [36], [37], [38].

In this work, we develop a compact optically-referenced all-PM Er:fiber OFC based on a silica optical-cube architecture. The 0.88-L optical comb module integrates the oscillator, amplifier, spectral-broadening stage, 780-nm frequency-doubling branch, and $f$-to-$2f$ interferometer. Without active temperature control, the free-running repetition frequency exhibits a peak-to-peak excursion of 75 Hz over 10 h, which corresponds to a relative variation of 0.375 ppm. Under the same laboratory conditions, this excursion is smaller than that of a mechanically mounted non-PM reference comb by a factor of 6.2. A CPT-based chip-scale atomic clock is further employed as the RF reference for $f_{CEO}$ and $f_{beat}$ locking loops. The resulting in-loop tracking instability, normalized to the optical carrier frequency, reaches $6.03\times10^{-17}$ at 1 s and $2.57\times10^{-18}$ at 10,000 s. These results demonstrate a compact and environmentally robust all-PM architecture for portable optical-frequency-comb systems.

## II. Experimental Setup and Principle

Figure 1 shows the schematic diagram of a fiber laser mode-locked by a Nonlinear Amplifying Loop Mirror (NALM). A 38-cm section of polarization-maintaining (PM) erbium-doped fiber (Er80-4) is employed as the gain medium. The remaining fiber sections, including the pigtails of all in-line components, consist of PM single-mode fiber (PM-SMF) with a total length of approximately 60 cm. The polarization beam splitters (PBSs), collimators, Faraday rotator (FR), and waveplate holders are directly bonded onto a silica substrate using optical-cube integration technology before fine optical alignment. This integration strategy significantly reduces the footprint of the free-space optical path. Moreover, by eliminating conventional mechanical mounts, the system effectively suppresses beam-pointing drift induced by thermal or mechanical perturbations.

Figure 2(a) presents photographs of the portable OFC module. The fiber optic path portion is fixed within grooves on an aluminum plate and rigidly secured by pouring encapsulation adhesive, which serves both to dissipate heat and to further reduce mechanical noise introduced by the fiber. The oscillator, amplifier, 780-nm frequency-doubling path, and $f$-to-$2f$ interferometer are all integrated into a single enclosure with a volume of only 0.88 L, and its physical size is 160 mm × 100 mm × 55 mm, as shown in Fig. 2(b).

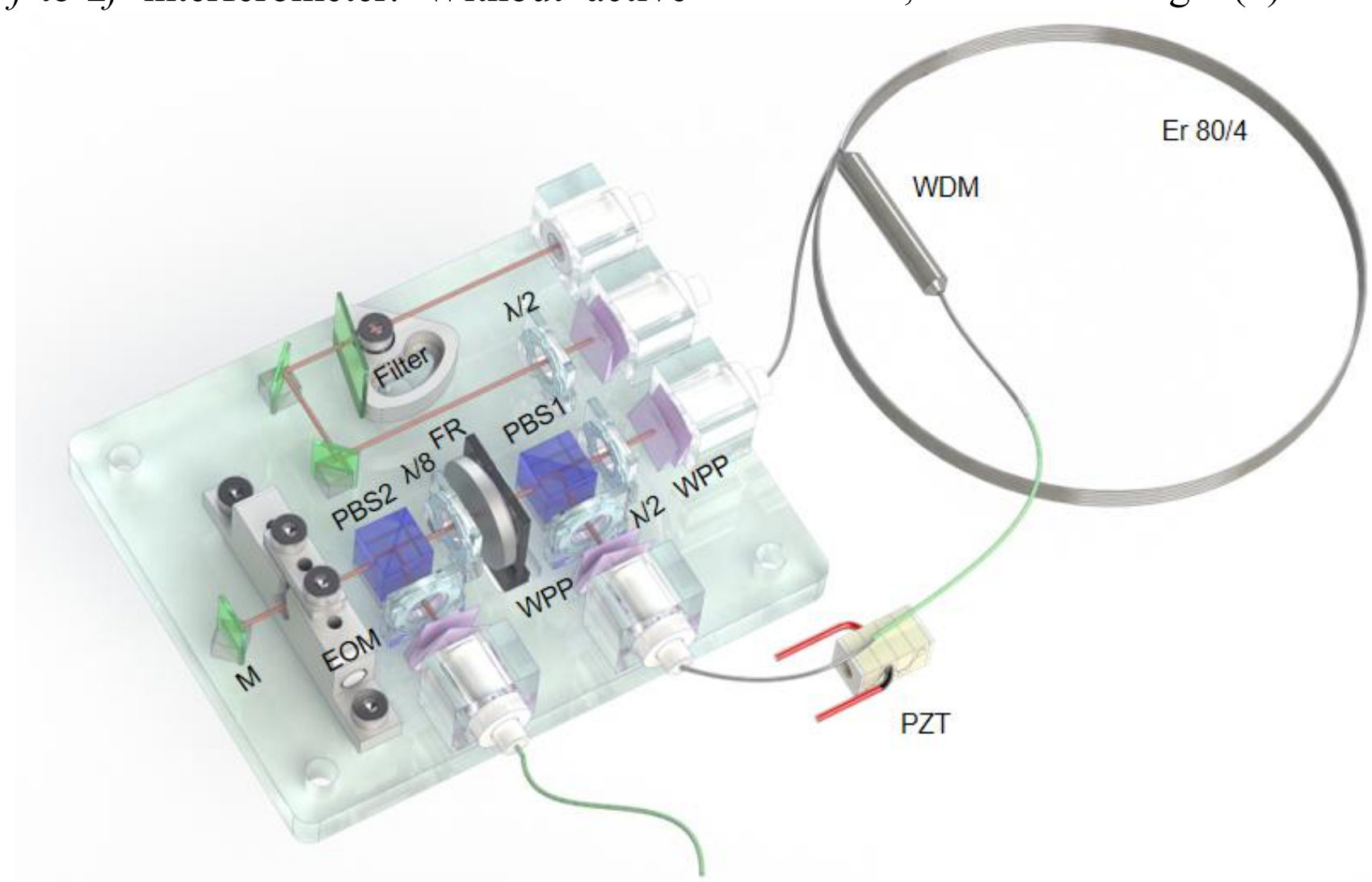

Fig. 1. Setup of the all-PM figure-9 fiber laser oscillator on Silica. λ/4, quarter-wave plate; λ/2, half wave plate, ISO, space isolator; PZT, piezoelectric transducer; WDM, wavelength division multiplexing; PBS, polarizing beam splitter; BS, beam splitter; PC, polarization controller; EOM, electro-optic modulator.

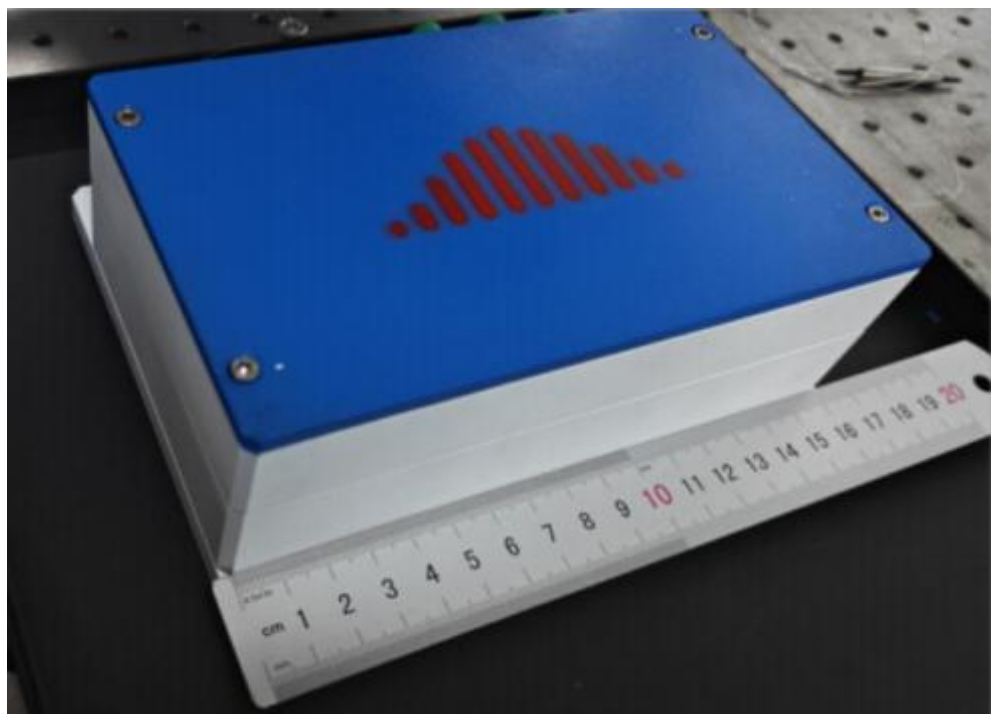

Fig. 2. Schematic external view of the portable optical frequency comb based on optical-cube integration.

### III. RESULTS

When the pump power is set to 1300 mW, a controllable nonreciprocal phase shift can be introduced between the orthogonal polarization components of the pulses by finely adjusting the 1/8 wave plate, thereby achieving mode-locking. Nevertheless, pulses tend to split under such high pump power. By slowly reducing the pump power to 1000 mW, stable single-pulse operation can be obtained, with an average output power of approximately 24 mW and an optical-to-optical conversion efficiency of 2.4%. At this operating state, fine tuning of the optical path for the linear arm of the resonant cavity can be realized by adjusting the position of the reflector, which barely disturbs the alignment of the original optical path and enables precise regulation of the repetition rate. In this experiment, the repetition rate is finally set to 200 MHz.

As illustrated in Fig. 3(a), the autocorrelation trace of the laser output pulse was measured using a commercial autocorrelator (Pulse Check USB, APE) with a pulse duration of 94 fs. The output spectral characteristics of the laser are measured by a optical spectrum analyzer (Yokogawa AQ6370D) at a spectral resolution of 0.05 nm. The 3 dB optical spectral bandwidth is up to 77.51 nm, as shown in Fig. 3(b) A broader spectral bandwidth is beneficial for generating Fourier-transform-limited pulses with shorter pulse duration.

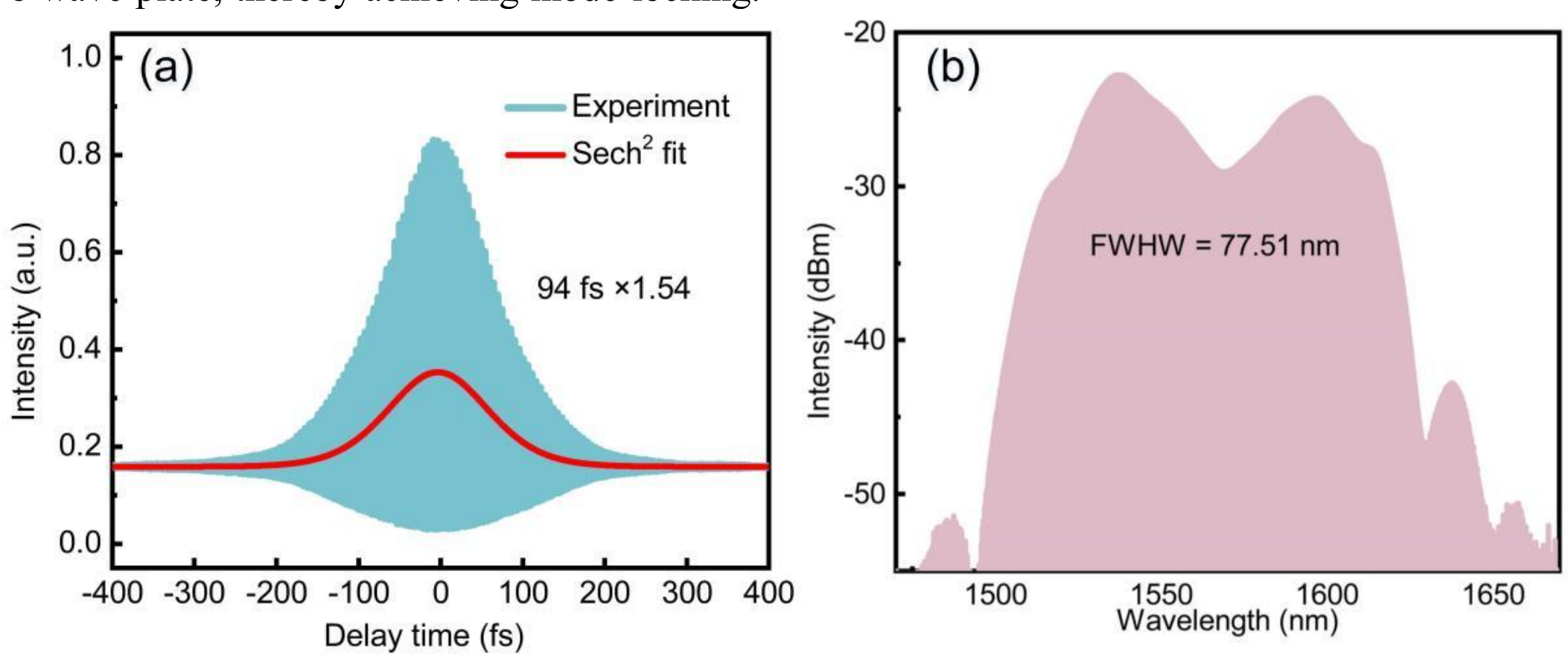

Fig. 3. All-PM figure-9 fiber laser oscillator on Silica: (a) Autocorrelation trace; (b) Optical spectrum.

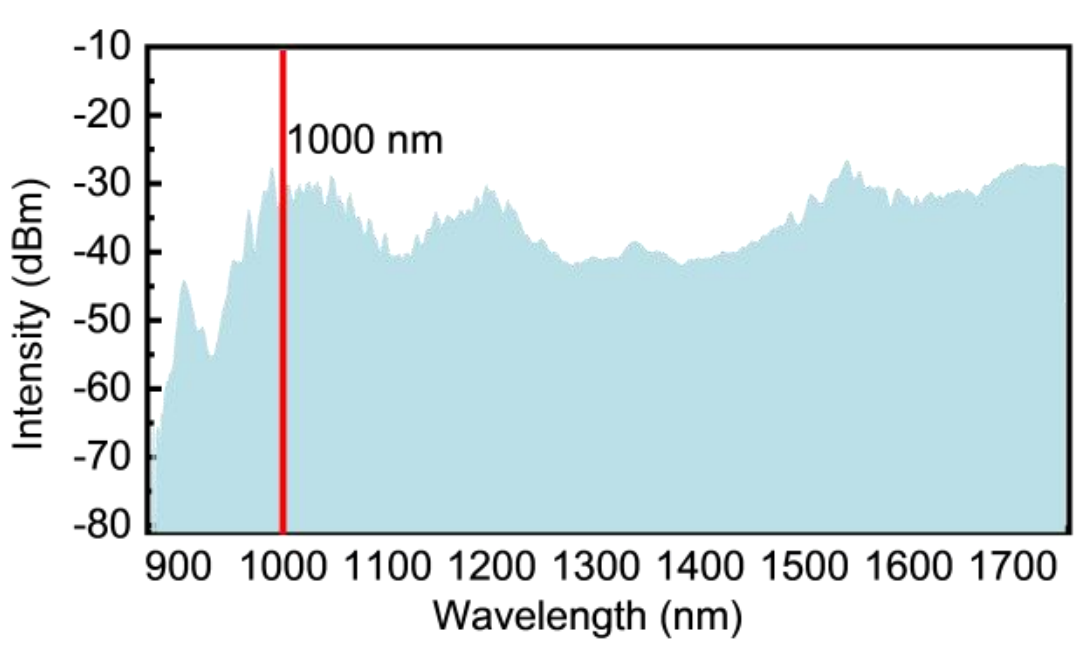

Fig. 4. Measured octave-spanning supercontinuum generated in polarization-maintaining highly nonlinear fiber (PM-HNLF).

To broaden the spectral coverage of the optical frequency comb, we adopt nonlinear fibers for spectral broadening. Meanwhile, polarization-maintaining highly nonlinear fibers (PMHN5 and PMHN) are employed to guarantee system stability. When the amplifier boosts the output power to 354 mW, the spectrum is extended to 1000 nm with a relatively flat profile, as depicted in Fig.4. Limited by the detection bandwidth of the optical spectrometer, the measurable spectral range only reaches 1700 nm.

To perform comparison with the optical comb based on the optical building block structure, a non-polarization-maintaining erbium-doped fiber optical comb is constructed on an aluminum plate via mechanical brackets. Figure 5 presents the 10-hour continuous measurement results of the repetition rate drift for the two optical combs under free-running operation. During the experiment, the lasers were placed in a standard laboratory environment without active temperature control, where the ambient temperature was maintained at 23±1 °C. The repetition rate data were collected using a commercial frequency counter (Keysight 53220A) with a gate time of 100 ms, and a commercial rubidium clock was adopted as the reference standard for the counter. Within the 10-hour test duration, the repetition rate drift of the photonic building-block optical comb was maintained within 75 Hz, corresponding to a relative repetition rate drift of 0.375 ppm. By contrast, the repetition rate drift of the optical comb with a traditional mechanical structure reached 650 Hz, with a relative drift of 2.31 ppm, representing a 6.2-fold difference in repetition rate drift between the two schemes [18]. Y. Feng et al. from Heriot-Watt University reported a relative repetition rate drift of 3.3 ppm for a solid-state mode-locked laser constructed on an aluminum baseplate [24]. Ruoao Yang et al. developed a Ytterbium-doped non-polarization-maintaining fiber laser based on a photonic building-block structure, achieving a relative repetition rate drift of 1.2 ppm [25]. Although this drift performance is greatly improved compared with that of the aluminum-based scheme, the drift level remains relatively high due to the adoption of non-polarization-maintaining fiber. This sufficiently demonstrates that the all-polarization-maintaining optical comb constructed with the photonic building-block structure exhibits superior stability. The low repetition rate drift indicates that the proposed OFC is insensitive to environmental variations and possesses high long-term stability, which facilitates long-term frequency locking.

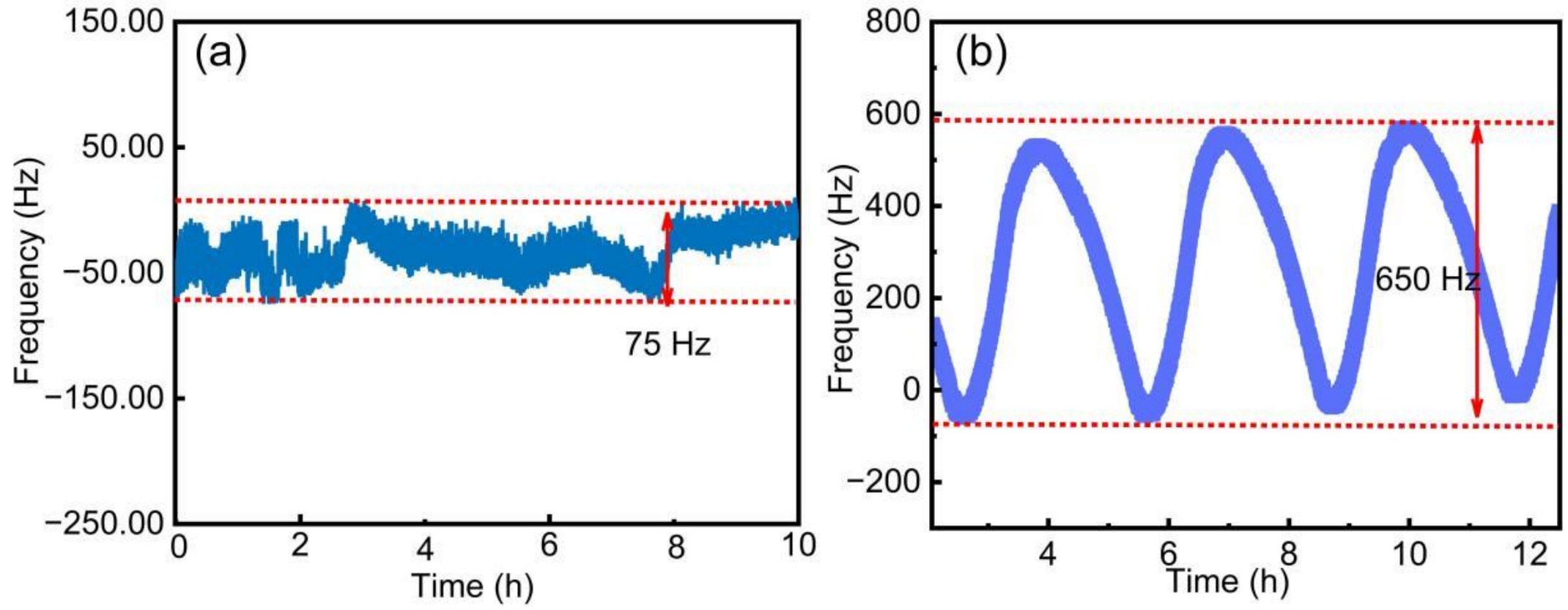

Fig. 5. (a) The free-running repetition-rate drift of the all-PM figure-9 fiber laser oscillator on Silica. (b) The free-running repetition-rate drift of the NPR-mode-locked fiber laser.

The pulse is amplified to 0.3 nJ by a nonlinear erbium-doped fiber amplifier, and the $f_{ceo}$ signal is subsequently extracted through an $f$-to-2$f$ waveguide module (Octave Photonics, COSMO). The signal-to-noise ratio of the free-running $f_{ceo}$ is estimated to be 36 dB at a resolution bandwidth of 100 kHz, as shown in Fig. 5(a). To achieve traceability to the optical frequency, the OFC is heterodyned with a home-made 780-nm optical frequency standard to generate the $f_{b780\ nm}$ signal. The signal-to-noise ratio of the free-running $f_{b780\ nm}$ reaches 32 dB, which provides sufficient margin for servo locking, as illustrated in Fig. 6(b).

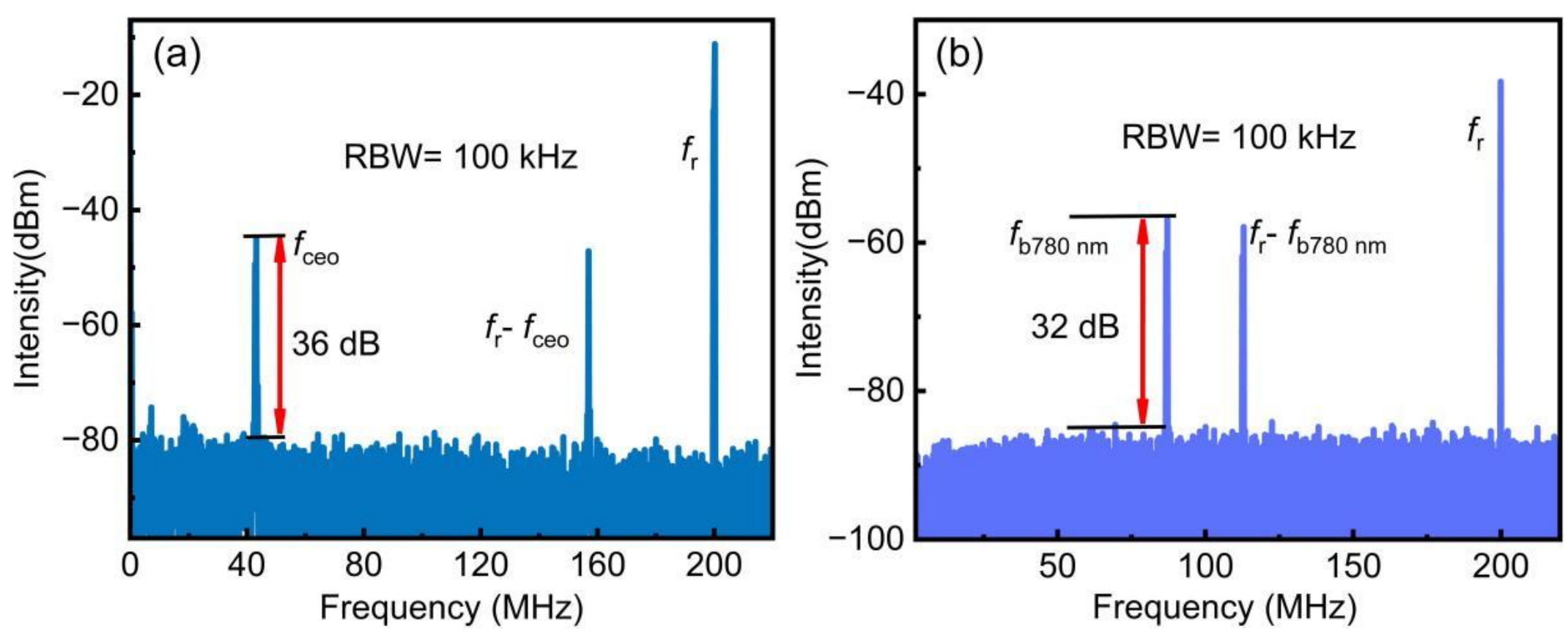

Fig. 6. (a) The frequency of free-running $f_{CEO}$. (b) The frequency of free-running $f_{b780\ nm}$.

Using a commercial locking circuit, the carrier-envelope offset frequency ($f_{CEO}$) is locked to a 100 MHz reference signal by adjusting the current. To improve the stability of $f_{CEO}$, various division ratios were tested in this section. A commercial rubidium atomic clock is adopted as the microwave reference for the OFC. As shown in Fig. 7(a), the servo bandwidth estimated from the servo response peak is 500 kHz, with a signal-to-noise ratio of 40 dB. Subsequently, the locked $f_{CEO}$ signal was subjected to frequency counting over 12 hours. To evaluate long-term performance, the $f_{CEO}$ was continuously recorded using Keysight 53230A frequency counters with a 1-s gate time, the frequency exhibits a peak-to-peak variation of 6 mHz, as illustrated in Fig. 7(b). The corresponding Allan deviation is presented in Fig 7(c); the frequency stability at 1 second is $2.98 \times 10^{-11}$ (normalized to the actual frequency). When $f_{CEO}$ is normalized to the optical frequency, the relative frequency stability at 1 second is $9.86 \times 10^{-18}$. Fig. 7(d) shows the phase noise after locking; the integrated phase noise of the locked $f_{CEO}$ (integration range: 1 Hz–10 MHz) is 436 mrad.

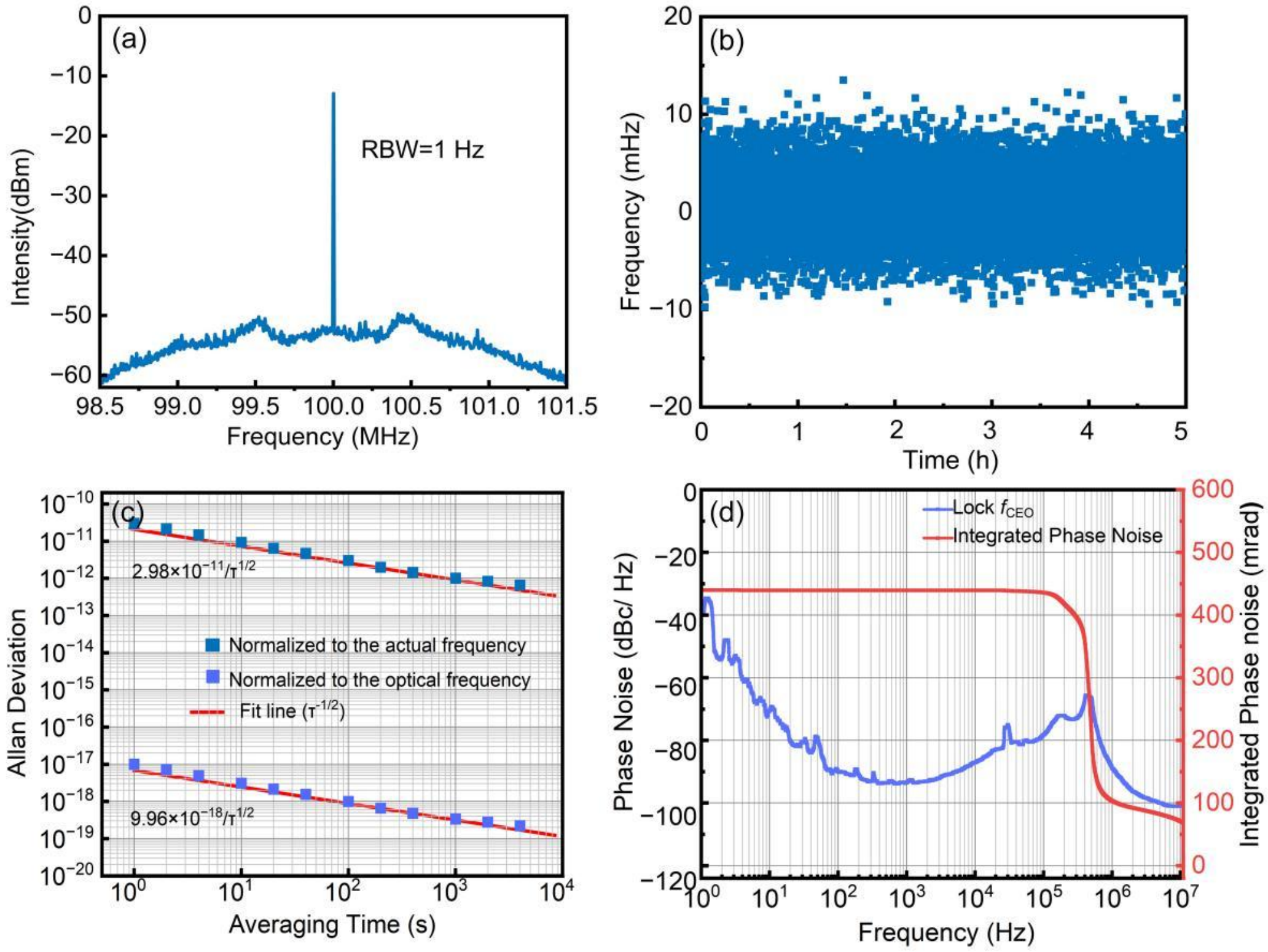

Fig. 7. Phase-locked $f_{CEO}$ signal: (a) RF spectrum; (b) Frequency jitter within 12 hours; (c) Frequency stability; (d) Power spectral density after phase locking $f_{CEO}$ signal (light blue curve), and integrated phase noise (red curve).

After obtaining the beat frequency signal, it is filtered and amplified to achieve an absolute power of 0 dBm. By optimizing the PID circuit parameters, the $f_{780\ nm}$ is locked to 100 MHz using both a PZT and an EOM. As shown in Fig. 8(a), the locked spectrum exhibits a gain peak, indicating that the locking bandwidth is 490 kHz (with a resolution bandwidth RBW=1 Hz). The frequency count is recorded by a counter, as shown in Fig. 8(b). After locking, the frequency stability of the $f_{780\ nm}$ signal reaches $2.75 \times 10^{-18}$ at 1s, as shown in Fig. 8(c). In theory, the Allan variance should decrease with a slope of 1/τ; however, limited by the counter, it follows a 1/τ slope. The phase noise of the locked $f_{780\ nm}$ nm signal is shown in Fig. 7(d). At an offset frequency of 1 Hz, the phase noise is −36 dBc/Hz. The integrated phase noise from 10 MHz down to 1 Hz is 136 mrad.

When the optical frequency comb is traced to an optical frequency reference, the following equation can be derived:

$$f_{rep} = (f_{laser} \pm f_{beat} \pm f_{CEO})/n \quad (1)$$

The $f_{beat}$ and $f_{CEO}$ signals are radio-frequency (RF) signals on the order of megahertz, while $f_{laser}$ corresponds to an optical frequency signal at hundreds of terahertz. The stability of $f_{rep}$ is primarily determined by the referenced optical frequency standard. Therefore, the stability of $f_{CEO}$ and $f_{beat}$ should be normalized to the optical frequency to characterize the tracking stability of the optical frequency comb against the optical frequency reference [27], [32].

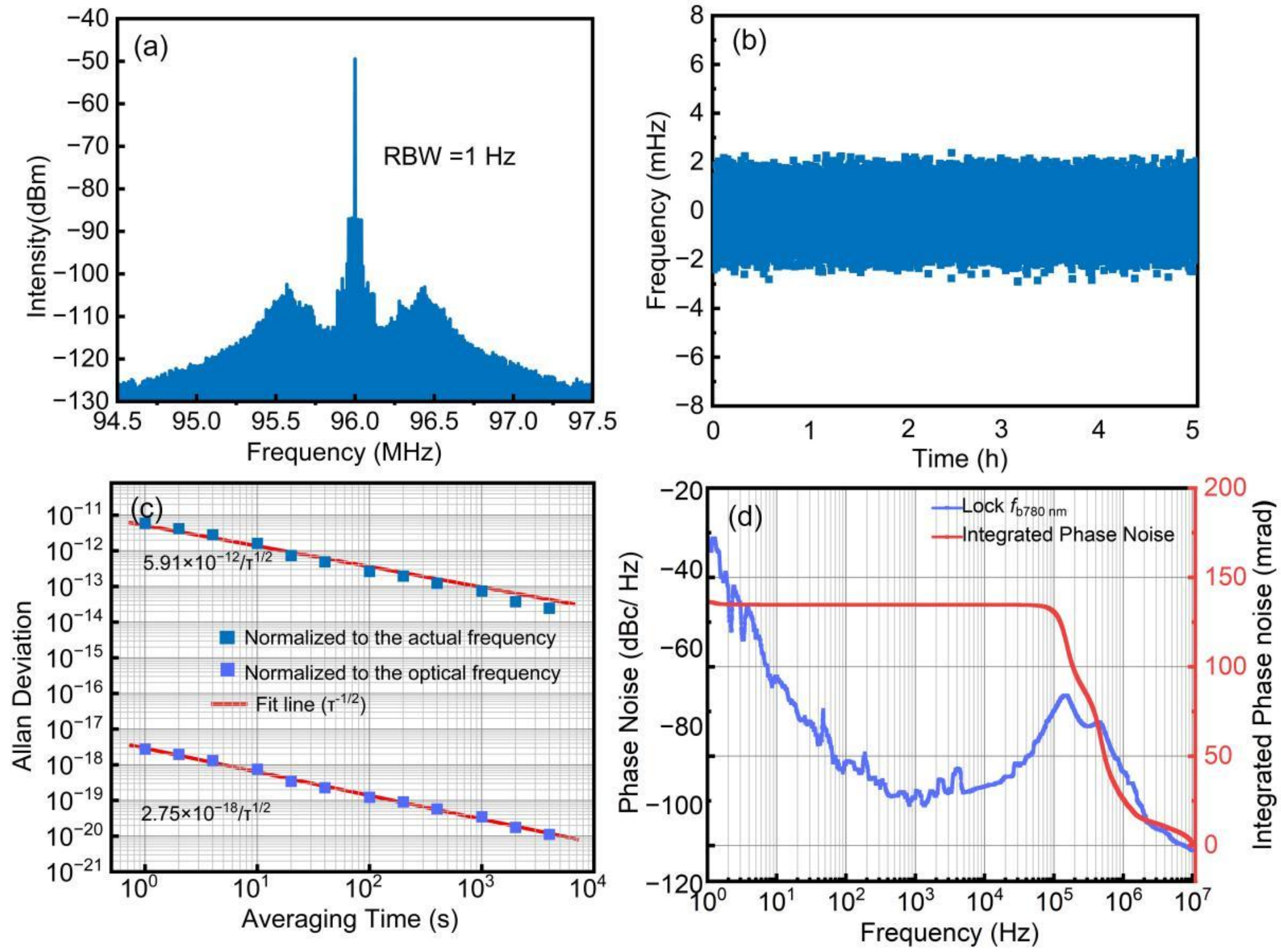

Fig. 8. Phase-locked $f_{b780\ nm}$ signal: (a) RF spectrum; (b) Frequency jitter within 12 hours; (c) Frequency stability; (d) Power spectral density after phase locking $f_{b780\ nm}$ signal (light blue curve), and integrated phase noise (red curve).

A CPT-based chip-scale atomic clock with a volume of 17.22 mL is employed as the common RF reference for the frequency synthesizers, phase-locking electronics, and frequency counters. The CSAC exhibits a measured fractional frequency instability of $6.99\times10^{-11}$at 1 s and $8.85\times10^{-12}$ at 10,000 s. These values characterize the microwave reference itself and are distinct from the optical-frequency-normalized in-loop tracking residuals of the stabilized comb.

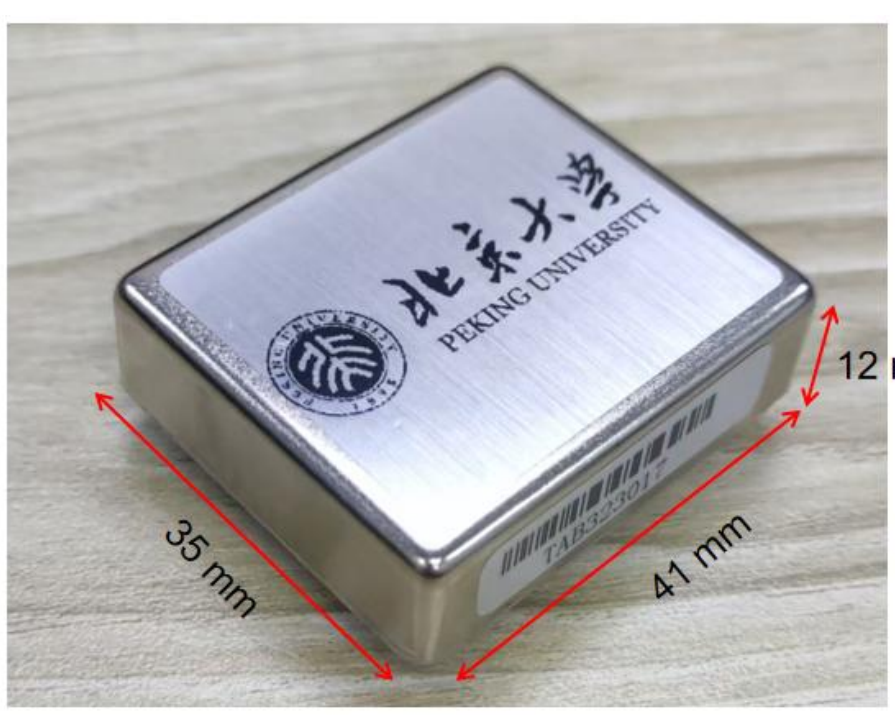

Fig. 9. Photograph of the chip-scale atomic clock.

Figure 10 illustrates the fractional frequency tracking instability of the OFC against the 780 nm optical frequency standard, when the $f_{CEO}$ and $f_{beat}$ loops are referenced either to the CPT-based CSAC or to a crystal oscillator. This tracking instability is obtained by combining the normalized in-loop residuals of the locked $f_{CEO}$ and $f_{beat}$ signals through Eq. (1); it characterizes the residual noise added by the locking system rather than the absolute stability of the RF reference itself, which explains why it lies far below the $10^{-11}$-$10^{-12}$ level of the CSAC (Fig. 8). With the crystal oscillator as the reference, the tracking instability is notably lower at intermediate times, reaching $5.90\times10^{-15}$ at 4000 s. With the CSAC reference, the tracking instability is $6.03\times10^{-17}$ at 1 s and $2.57\times10^{-18}$ at 10,000 s. Such tracking performance enables the OFC to support frequency downconversion, frequency comparison, and high-precision metrology in portable optical clock systems.

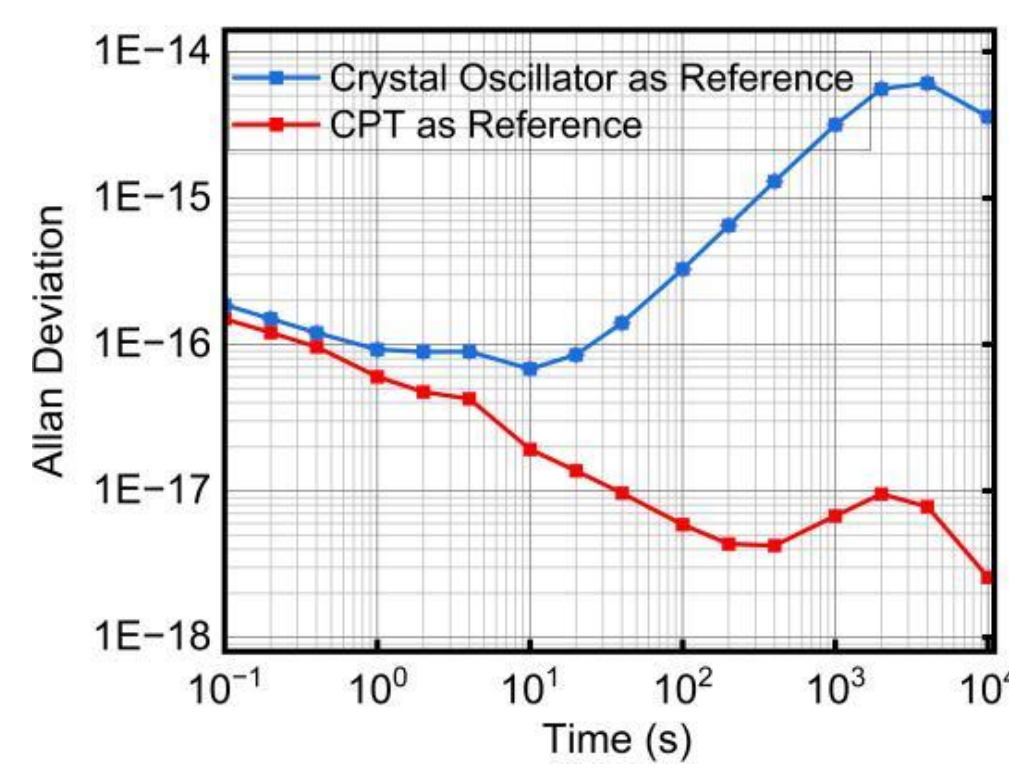

Fig. 10. Tracking stability of the optical frequency comb. (the blue trace corresponds to the optical frequency comb referenced to a crystal oscillator, while the red trace represents the comb referenced to the chip-scale atomic clock.)

## IV. Conclusion

We have demonstrated a compact optically-referenced all-polarization-maintaining Er:fiber optical frequency comb based on a silica optical-cube architecture. The optical comb module, with a volume of 0.88 L, integrates the oscillator, amplifier, spectral-broadening stage, 780-nm frequency-doubling branch, and *f*-to-2*f* interferometer. In the absence of active temperature control, the free-running repetition frequency exhibits a peak-to-peak excursion of 75 Hz over 10 h, corresponding to a relative variation of 0.375 ppm. Under the same laboratory conditions, this excursion is 6.2 times smaller than that of a mechanically mounted non-PM reference comb. A CPT-based chip-scale atomic clock is further employed as the common RF reference for the comb-locking electronics, thereby reducing the footprint of the microwave-reference subsystem. The resulting in-loop tracking instability, normalized to the optical carrier frequency, reaches $6.03\times10^{-17}$ at 1 s and $2.57\times10^{-18}$ at 10,000 s. The demonstrated architecture provides a compact and technically scalable platform for the further development of portable optical-frequency-comb systems. Future work will focus on independent out-of-loop verification and controlled thermal and vibration tests under field-relevant conditions.